\documentstyle[prl,aps,amsfonts,amssymb,twocolumn,epsfig]{revtex}
\begin{document}
\draft
 \wideabs{

\title{Probing the electron-electron interaction in a diffusive gold wire
using a controllable Josephson junction  }

\author{J.J.A.Baselmans, B.J. van Wees, and T.M. Klapwijk
$^\ast$}
\address{Department of Applied Physics and Materials Science Center, University
of Groningen,\\ Nijenborg 4, 9747 AG Groningen, The Netherlands \\
$^\ast$Department of Applied Physics (DIMES), Delft University of
Technology, \\ Lorentzweg 1, 2628 CJ Delft, The Netherlands \\}
\date{\today}

\maketitle

\begin{abstract}
We have studied the critical current of a diffusive superconductor -
normal metal - superconductor (SNS) Josephson junction as a function of
the electron energy distribution in the normal region. This was realized
in a 4 terminal device, in which a mesoscopic gold wire between two
electron reservoirs is coupled in its center to two superconducting
electrodes. By varying the length of the wire and applying a voltage over
it we are able to control the electron distribution function in the center
of the wire, which forms the normal region of the SNS junction. The
observed voltage and temperature dependence are in good agreement with the
existing theory on diffusive SNS junctions, except for low energies.
However, an electron-electron interaction time $\tau_0$ =10 ps was found,
which is three orders of magnitude faster than expected from theory.

\end{abstract}
}

The study of electron-electron interaction in metals has attracted
considerable attention recently. One of the reasons is an experiment
performed by Pothier et al. \cite{Pothier} in which the electron
distribution function in a mesoscopic wire connected to large electron
reservoirs was measured using a superconducting tunnel junction. The shape
of the distribution function depends, at low temperatures, on the
effective electron-electron interaction experienced by the electrons, and
the voltage $V_c$ applied over the reservoirs. If the wire is short, the
electrons will keep their energy while traversing the wire, and the
distribution function will show a double step structure, with a separation
between the steps of $eV_c$. If the wire is much longer, so that the
diffusion time $\tau_D$ through the wire strongly exceeds the
electron-electron interaction time $\tau_0$, then a local thermal
equilibrium of the electron system will be regained with an effective
temperature $T_{eff}$, depending on $eV_c$. The electron interaction time
constant $\tau_0$ obtained from these experiments yields, assuming a local
two particle interaction, $\tau_0 \approx$ 1 ns. for copper, which is two
orders of magnitudes faster than expected from theory
\cite{Schmid,Altshuler1,Altshuler2}.\\

\begin{figure}[b]
\centerline{\psfig{figure=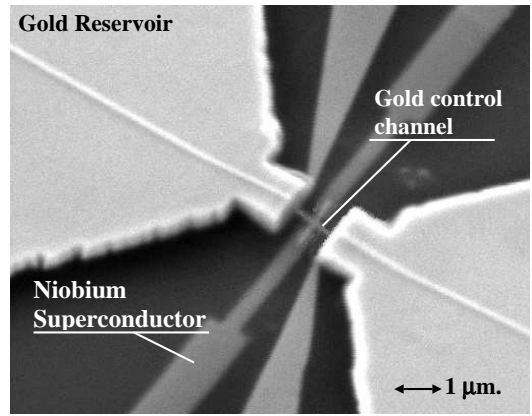,width=7 cm,clip=}}
\vspace{0.3cm} \caption{One of the SNS devices. In this case the length of
the control channel is 1 $\mu$m, The gold control channel (100 nm wide)
shows a cross in its center, of which the two arms are covered by niobium
electrodes, thus forming the SNS junction in the center of the channel.
Contacts at the channel ends measure the control voltage.} \label{Device}
\end{figure}

\begin{figure}[tbp]
\centerline{\psfig{figure=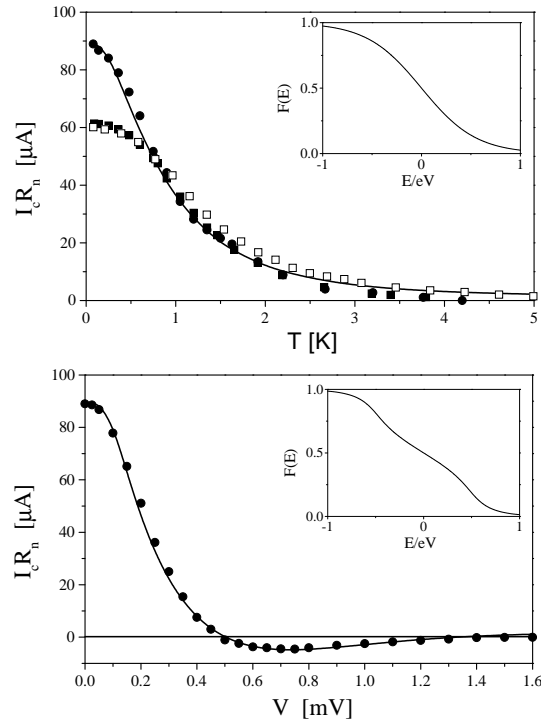,width=7 cm,clip=}}       
\caption{$I_cR_n$ as a function of temperature and voltage for both
devices. The top panel shows the temperature dependence of device 1
(circles) and device 2 (filled squares), and the voltage dependence of
device 2 (open squares) expressed in units of $T_{eff}$. The bottom shows
the voltage dependence of device 1. The best fits to the theory are given
by the solid lines, assuming a distribution function as shown in the
insets.}
\label{results}
\end{figure}

\begin{figure}[tbp]
\centerline{\psfig{figure=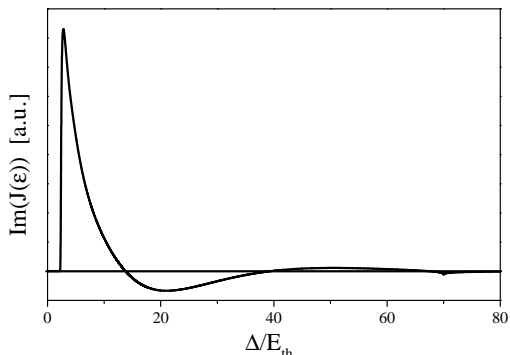,width=7 cm,clip=}}
\caption{Supercurrent carrying density of states, using
$\Delta$/$E_{th}$=70, $\Delta$=1.5 meV (Niobium) and $E_{th}$=21$\mu$eV.
 (from ref. [7,8])} \label{scdos}
\end{figure}

\begin{figure}[tbp]
\centerline{\psfig{figure=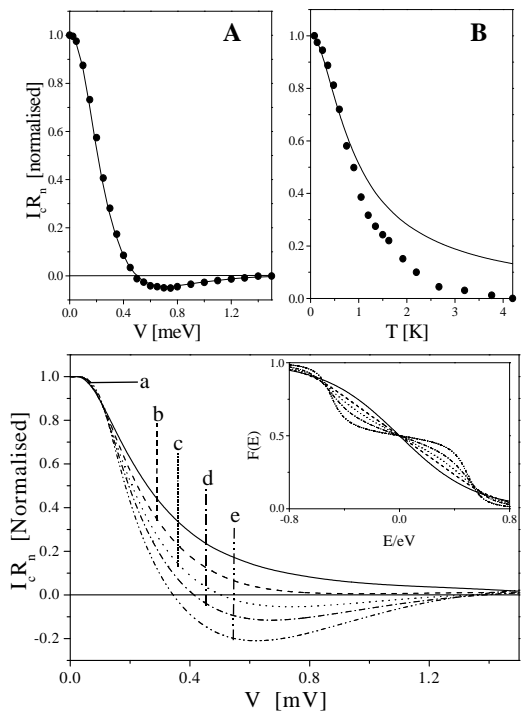,width=7 cm,clip=}}       
  \caption{
The two top figures show a theoretical fit to both the voltage and
temperature dependence of device 2, assuming a perfect step distribution
function and, as a result, smaller negative contributions in
Im(J($\epsilon)$). The bottom figure shows the strong dependence of the
$I_cR_n$ vs. $V_c$ behaviour on the effective electron-electron
interaction in the control channel, $\tau_0$/$\tau_D$. The inset shows the
corresponding distribution functions.
  $\tau_0/\tau_{D}$=:
  $\bf{a}$: 0, $\bf{ b}$ 0.1 $\bf{c}$ 0.2 $\bf{d}$ 0.4 $\bf{e}$ 1.
} \label{relaxation}
\end{figure}

On the other hand, it is in principle also possible to use a diffusive
superconductor - normal metal - superconductor junction as a probe to
study electron-electron interaction effects in a mesoscopic wire. The
prediction is that $I_cR_n$-product (critical current times normal state
resistance) of such a SNS junction is very sensitive to the exact electron
distribution function \cite{Volkov1,Volkov2,Wilhelm,Yip}. In this Letter
we report experiments in which we studied the critical current of a SNS
junction as a function of the electron distribution. This was realized
using the device shown in Fig. \ref{Device}. A diffusive gold wire
(control channel) between two very large electron reservoirs is coupled to
two niobium electrodes by means of a cross shaped extension in the center
of the wire. We study two different device geometries, which differ only
in the length of the control channel. We made 3 devices with a short
control channel, $L_{control}$ = 1 $\mu$m, and two devices with a long
control channel, $L_{control}$ = 9 $\mu$m. The behaviour of identical
devices was similar, we therefore present only two devices in the
remainder of the text: Device 1, with a short control channel and device 2
with a long control channel.
The devices are made by means of e-beam lithography using a double layer
of PMMA resist and a subsequent lift-off process on top of a thermally
oxidised Si wafer. The 100 nm wide control channel is evaporated first,
which consists of 40 nm of gold on top of 5 nm Ti, which was used to
improve adhesion. Subsequently 70 nm of niobium was sputter deposited
after an in-situ argon etching of the gold contacts to ensure a high
interface transparency. In a last step the very thick (475 nm) gold
reservoirs were deposited. The size of the reservoirs is in the order of a
millimeter because they should also act as effective cooling fins to
prevent unwanted electron heating \cite{UrbinaClarke,Henny}. The gold has
a diffusion coefficient of 0.020 $m^2$/s which results in an estimated
diffusion time through the channel of $\tau_D = 50$ ps. and 4 ns. for
device 1 ($L_{control}= 1 {\mu}$m) and 2 ($L_{control}= 9 {\mu}$m)
respectively. The SNS junctions all have a normal state resistance of 2.1
$\Omega$, and a separation of the niobium electrodes of 375 nm.
\\
In the experiment we measured the current-voltage (I-V) characteristics of
the SNS junction as a function of both the voltage $V_c$ applied over the
control channel as well as the bath temperature $T_B$. In the first case
the bath temperature is kept at 100 mK to obtain a very sharp electron
distribution function in the reservoirs. RC filtering at room temperature
and copper powder filtering at the bath temperature are used in all
measurements to reduce external noise and hence, unintentional heating of
the electrons. From these measurements we obtained the critical current.
In Fig \ref{results} we show the results. The voltage dependence of device
1, as shown in the bottom panel of the figure, shows the transition to a
$\pi$-junction at control voltages $V_c$ $>$ 0.5 meV, similar to previous
experiments \cite{me}. However, the maximum supercurrent in the
$\pi$-state is much smaller than expected when assuming a perfect step
distribution function. The top panel of Fig. \ref{results} shows the
results obtained in the limit of a thermal distribution function: The
filled circles and squares represent the temperature dependence of the
$I_cR_n$ product of device 1 and 2 respectively. The empty squares
represent the behaviour of the critical current of device 2 as a function
of $V_c$, expressed in terms of the effective temperature $T_{eff}$,
assuming a perfect thermalisation of the electrons in the wire
\cite{commthermal}. The effective electron temperature in the center of
the control channel can now be calculated using the Wiedemann-Franz law
\cite{Pothier}:
\begin{equation}
    T_{eff}~=~\sqrt{{{T_B}^2+(a{\cdot}V)^2}}
\label{Teff}
\end{equation}
with a=3.2 K/mV. All curves show the expected monotonic decrease in $I_c$
with increased temperature \cite{Alberto}, and at higher temperatures all
curves lay essentially on top of each other, with the exception that
$I_cR_n$ vs. $T_{eff}$ (open squares) is somewhat larger than $I_cR_n$ vs.
$T_B$ (solid squares) at higher temperatures. This indicates that
$T_{eff}$ increases slower with increasing $V_c$ than expected, which is
probably due to the onset of electron - phonon interactions at higher
electron energies \cite{commep}. This leads to an extra cooling of the
electrons, and thus to a larger supercurrent. A very striking difference
between both devices, which is counterintuitive and seems to be in
disagreement with theoretical predictions \cite{Yip}, arises at low
temperatures, T $<$ 800 mK.: device 1, with the lowest control channel
length, and thus the strongest coupling to the normal reservoirs, shows a
much higher $I_cR_n$-product (87 $\mu$eV), than device 2 ($I_cR_n$=56
$\mu$eV), whereas one would expect a reduction in supercurrent with
increased coupling to a normal reservoir. This observation is consistent
with a previous experiment, which yielded an even higher $I_cR_n$-product
for a system with an even stronger coupling to the normal reservoirs
\cite{me}.
The next step in the analysis is to compare the data to the existing
theory on diffusive SNS junctions. This theory predicts that supercurrent
will be carried by a supercurrent carrying density of states,
Im(J($\epsilon$)), which can be calculated using the quasi-classical
Green's function theory \cite{Volkov1,Volkov2,Wilhelm,Yip}. The positive
and negative parts of the supercurrent carrying density of states, as
shown in of Fig. \ref{scdos}, represent, at a given phase difference
 $\pi/2$, energy dependent contributions to the supercurrent in the positive
and negative direction. The critical supercurrent now depends strongly on
the occupation of this continuum of states, and thus on the electron
distribution function in the normal region, and can be calculated
according to \cite{Volkov1,Volkov2,Wilhelm,Yip}:
\begin{equation}
I_{c}R_{N}={\int_{-\infty}^{\infty}}\partial\epsilon[1-2f(\epsilon)]Im(J(\epsilon))
\label{IcRn}
\end{equation}
where ImJ(($\epsilon$) is the supercurrent carrying density of states and
f($\epsilon$) the electron distribution function. The transition to a
$\pi$-junction in case of a perfect step function  can be understood from
this, for in this case the distribution function will be exactly 0.5 over
a region e$V_c$ around the Fermi energy, resulting in a zero contribution
to the supercurrent over this energy range. At a large enough value of
$V_c$, its magnitude depending on $E_{th}$, all positive contributions to
the supercurrent will be blocked, which obviously changes the direction of
the supercurrent.
We tried to fit the temperature and voltage dependence of the
$I_cR_n$-product, using eqn. \ref{IcRn}, in which the Thouless energy as
well as the exact shape of the distribution function were used as fit
parameters. However, the shape of the distribution function can be
calculated as a function of the effective electron-electron interaction
time in the control channel, which leaves only $E_{th}$ and $\tau_0$ as
fit parameters. \\
\\
The calculation of the distribution function was performed using the
method described in \cite{Pothier}, assuming a local two particle
interaction with an interaction kernel given by $\frac{1}{\tau_0
\epsilon^2}$, with $\epsilon$ the transferred energy and $\tau_0$ the
electron-electron interaction time constant. In this model the shape of
the  distribution function depends on 2 parameters: The ratio of the
electron interaction time with the electron diffusion time,
${\tau_0}/{\tau_D}$ and on ${kT_e}/{eV_c}$, which is the ratio of the
electron temperature in the reservoirs $T_e$ and the control voltage
applied. $T_e$ in the reservoirs cannot be taken as a constant and equal
to $T_B$, because the electron heat conductance of the reservoirs is
finite and electron-phonon interaction is strongly reduced at low
temperatures. These arguments combined with the fact that rather large
voltages are used in these experiments, make it necessary that the heating
of the electrons in the reservoirs is calculated explicitly. This was done
using a model presented by Henny et al. \cite{UrbinaClarke,Henny}. The
result is that, at $T_B$ $\approx$ 100 mK, $T_e$ increases with increasing
control voltages, but this increase is linear if $V_c$ $>$ 0.4 mV. In this
limit $kT_e$/$eV_c$ is constant, which implies that the shape of the
distribution function is also constant. Higher values of the applied
voltage merely change the energy range of the distribution function, not
its shape \cite{Pothier}.
\\
The results of the fits are shown by the solid lines in Fig.
\ref{results}, in which the maximum value of the $I_cR_n$-product of
device 1 was used as normalisation constant. The inset shows the
distribution functions (in units of $E/eV_c$) in the limit where $kT/eV_c$
of the reservoirs is constant. We found a value of the Thouless energy
given by $E_{th}$=21 meV, which is in good agreement with the expected
$I_cR_n$-product of a diffusive SNS junction, $\frac{I_cR_n}{E_{Th}}= \pi$
\cite{Wilhelmold}, as well as the value reported in \cite{Yip} for the
cross geometry discussed here, $\frac{I_cR_n}{E_{Th}}= 5$. The effective
Nb electrode separation using this value of $E_{th}$ is l=800nm, which is
in between the minimum Nb separation and the maximum extent of the gold
under the Nb. It is clear from Fig \ref{results}that the agreement between
the experiment and the calculations is excellent, apart from the low
voltage region of device 2 discussed previously. However, from the insets
is is also clear that the shape of the distribution functions indicates a
significant electron-electron interaction: The distribution function has a
Fermi-Dirac shape in the case of device 2, which was expected, but it is
also very rounded already in the case of device 1. This strong rounding is
responsible for the small magnitude of the supercurrent in the $\pi$
state. The interaction time constant $\tau_0$ obtained from this analysis
is device independent, and given by $\tau_0$ $\approx$ 10 ps. We have
found that choosing a larger relaxation time constant in combination with
another normalisation constant or another value of $E_{th}$ does not yield
reasonable and consistent fits for both devices. Moreover, the dependence
of the $I_cR_n$-product on the control voltage is extremely sensitive to
the effective electron relaxation in the control channel,
$\tau_D$/$\tau_0$, and thus on the exact shape of the distribution
function, as shown at the bottom of Fig. \ref{relaxation}. Here the
calculated distribution function as well as the resulting $I_cR_n$ - $V_c$
behaviour is plotted for 5 different distribution functions, using five
different ratio's of $\tau_0/\tau_D$. It is clear that even small
deviations from a thermal distribution function (curve b) results in a
quite strong difference in the voltage dependence of the $I_cR_n$-product,
even leading to the transition to a $\pi$-junction at relatively rounded
distribution functions (curve c).

Up till now we have based our analysis on the assumption that
Im(J($\epsilon$)) is correct. However, the small values of the $\pi$
supercurrent could in principle also be explained assuming a perfect step
distribution function if the negative contributions to the supercurrent of
$Im(J(\epsilon))$ would be smaller. This assumption yields the fit of the
$I_cR_n$ vs. $V_c$ data of device 1 as shown in panel A of Fig.
\ref{relaxation}. However, if this form of $Im(J(\epsilon))$ is used to
calculate the temperature behaviour, we get a strong disagreement between
measurements and theory, as shown in panel B. We therefore conclude that
the theory which describes the supercurrent as a function of the
distribution function is in principle correct, except at low energies, for
it fails to predict the difference in $I_cR_n$-product between the devices
at low temperatures.
On the other hand, the value of $\tau_0$ implies an upper bound for the
quasiparticle lifetime, valid over the energy range of these experiments
(0.01 - 1.8 meV), given by $\tau$(E) $<$ $\tau_0$/ln(E/0.01 meV)
\cite{Pothier}. This yields a maximum quasiparticle lifetime, at E=0.01
meV, given by $\tau_{e-e}$ $\sim$ 10 ps. However, the electron out
sacttering time can be calculated using the existing theory on
electron-electron interactions in a diffusive 1D wire, according to
\cite{Schmid,Altshuler1,Altshuler2}:
\begin{equation}
\tau_{ee}(E)=\sqrt{2}\hbar n_0 S \sqrt{\frac{\hbar D}{E}}
 \label{alt}
\end{equation}
with $n_0$ the density of states around the Fermi level, and S the area
cross section of the wire. However, this yields, using
$n_0$=1.9$\cdot10^{28}$ $m^{-3}eV^{-1}$ $\tau_{ee}$(0.01 meV)=80 ns, a
difference of three orders of magnitude with our result. We note that
experiments on simular wires using superconducting tunnel junctions
\cite{Pothier,Pothier2} give a a value of $\tau_0$ 1 ns and 0.1 ns for
copper and gold respectively. The difference between our data and these
measurements can be explained by the fact that we used a thin Ti adhesion
layer (which has a very fast phase relaxation at low temperatures, caused
by electron-electron interaction \cite{Ti}). Alternatively the strong
coupling between the niobium and the gold in the junction might influence
the electron interaction. However, the difference with the existing theory
\cite{Schmid,Altshuler1,Altshuler2} is more than 3 orders of magnitude,
which can hardly be explained using such arguments. Very recently
\cite{Altnew1} the experimental procedure of Ref. \cite{Pothier} to obtain
the electron distribution was questioned, as well as the method discussed
here \cite{Altnew2}. The prediction is that the shot noise in the control
channel associated with the non-equilibrium distribution function yields a
smearing of the density of states in the superconducting electrodes,
resulting in a supercurrent reduction. Our observations are not in
agreement with the predicted \cite{Altnew2} behaviour of $I_c$ vs. $V_c$
(see bottom panel of Fig. \ref{results}). Also, assuming the theoretical
value $T_0$ $\approx$ 80 ns would result in a transition to a $\pi$-state
of device 2 as well, which we did not observe in the experiment.
\\
We greatly acknowledge H. Pothier,  for helpful discussions and the
computer program to calculate the distribution functions, and F. Wilhelm
for the computer program to calculated the supercurrent carrying density
of states. Furthermore we wish to thank A. Morpurgo for his initiating
role leading to this work. This work was supported by the Nederlandse
Organisatie voor Wetenschappelijk Onderzoek (NWO) through the Stichting
voor Fundamenteel Onderzoek der Materie (FOM).


\begin{references}
\bibitem{commthermal} The assumption of a thermalised distribution
function, e.g. $\tau_{e-e}\ll\tau_D$ is correct in this case, as will be
shown in the remainder of the text.
\end{references}
\end{document}